\documentclass[iicol,pdflatex,sn-basic]{sn-jnl}

\usepackage{graphicx}%
\usepackage{multirow}%
\usepackage{amsmath,amssymb,amsfonts}%
\usepackage{amsthm}%
\usepackage{mathrsfs}%
\usepackage[title]{appendix}%
\usepackage{xcolor}%
\usepackage{textcomp}%
\usepackage{manyfoot}%
\usepackage{booktabs}%
\usepackage{algorithm}%
\usepackage{algorithmicx}%
\usepackage{algpseudocode}%
\usepackage{listings}%

\theoremstyle{thmstyleone}%
\theoremstyle{thmstyletwo}%

\theoremstyle{thmstylethree}%

\begin{document}

\title[Photospheric abundances of Altair]{Photospheric abundances of the rapidly-rotating 
A-type star Altair}


\author[]{\fnm{Yoichi} \sur{Takeda}}\email{ytakeda@js2.so-net.ne.jp}

\affil[]{11-2 Enomachi, Naka-ku, Hiroshima-city, \postcode{730-0851}, \country{Japan}}


\abstract{
Altair is an A-type star known to have an appreciably oblate shape owing to 
its very fast rotation ($\sim 300$\,km\,s$^{-1}$). Despite of numerous 
publications on this star, its chemical abundances have been scarcely 
investigated so far, presumably because of the practical difficulty that 
spectral lines are considerably broadened by rapid rotation and badly blended 
with each other. Motivated by this situation, a spectroscopic analysis 
was conducted to study the photospheric abundances of Altair by using
the synthetic spectrum-fitting technique, in order to clarify whether 
or not any chemical peculiarities exist. The microturbulent velocity
was determined to be 2.9\,($\pm 0.9$)\,km\,s$^{-1}$ by requiring that 
the metallicity does not show any systematic region-dependence.
Then, the abundances of 17 elements (C, N, O, Mg, Al, Si, S, Ca, Sc, Ti,
Cr, Mn, Fe, Ni, Zn, Sr, Ba) were derived, where the non-LTE effect was
taken into consideration as much as possible. The results revealed 
considerable region-by-region dispersion (several tenths dex or even more), 
reflecting the difficulty of reliable abundance determination for such a very 
rapid rotator. Nevertheless, the differential mean abundances relative to 
the Sun turned out to fall within $-0.5 \lesssim$\,[X/H]\,$\lesssim +0.3$ 
for all elements without any dependence upon the atomic number. Accordingly,
we may conclude that (1) no appreciable anomalies of chemical abundance 
patterns exist in the atmosphere of Altair, (2) but its global metallicity 
is likely to be slightly subsolar ($\sim -0.2$\,dex on the average). 
}




\maketitle

\section{Introduction}\label{sec1}

Altair (= $\alpha$~Aql = HR~7557 = HD~187642 = HIP~97649) is 
an A-type main-sequence star (spectral type: A7~V), which is 
easily recognizable as a first-magnitude star ($V = +0.76$~mag) 
near to the Milky Way in the sky of summer night.

It is notable that this star is spectroscopically known to rotate 
very rapidly ($v_{\rm e}\sin i \sim$\,200--250\,km\,s$^{-1}$, where 
$v_{\rm e}$ is the equatorial rotational velocity and $i$ is the 
inclination angle of rotational axis). Although this is not uncommon 
in A-type stars, the remarkable feature of Altair is that the effect
of rapid rotation (e.g., flattening effect or gravitational darkening) 
is directly measurable by interferometric observations thanks to 
its proximity from us (distance of only 5.14~pc according to the 
{\it Hipparcos} parallax). Actually, the parameters relevant to such a 
flattened rotator (e.g., effective temperature $T_{\rm eff}$ and radius $R$ at 
the pole and equator, etc.) as well as $v_{\rm e}$ and $i$ have been separately
determined interferometrically, as summarized in Table~\ref{tab1}.
As such, the physical condition over the surface of Altair is rather well
understood nowadays.   
   
\setcounter{table}{0}
\begin{table*}[h]
\caption{Altair's parameters determined by interferometric observations.}
\scriptsize
\begin{center}
\begin{tabular}{ccccccccccl} 
\hline\hline
Literature &  $M$  & $T_{\rm eff,p}$ & $T_{\rm eff,e}$ & $R_{\rm p}$ &  $R_{\rm e}$ & 
 $v_{\rm e}$ & $i$ & $v_{\rm e}\sin i$ & $\beta$ & Remark \\
(1) & (2) & (3) & (4) & (5) & (6) & (7) & (8) & (9) & (10) & (11) \\
\hline
\citet{vanBelle_etal:2001}&  & 7680 &   & 1.887 &   &   &   & 210 &  &  \\
\citet{Ohishi_etal:2004}    &  & 7750 &   &       &   &   &   &     &  &  \\
\citet{DomicianodeSouza_etal:2005}  & 1.8 & 8500 & 6509 & 1.711 & 2.117 & 277 & 55 & 227 & 0.25 &  \\
\citet{Peterson_etal:2006} &  & 8740 & 6890 & 1.636 & 1.988 & 273 & 64 &  245 & 0.25 &  \\
\citet{Monnier_etal:2007}  & 1.791 & 8450 & 6860 & 1.634 & 2.029 & 285 & 57.2 & 240 & 0.19 & adopted here\\ 
\citet{Hadjara_etal:2014}  & 1.791 & 8912 & 7372 & 1.639 & 2.000 & 269 & 57.3 & 226 & 0.19 & \\
\citet{Bouchaud_etal:2020} & 1.86  & 8621 & 6780 & 1.565 & 2.008 & 313 & 50.65 & 242 & 0.185 & \\
\hline
input/output of SEDINT & 1.8 &  8450 & 6884 & 1.63  & 2.02  & 285 & 57 & 239 & 0.19 & \\
\hline
\end{tabular}
\end{center}
(1) Reference. (2) Mass (in unit of the solar mass $M_{\odot}$). (3) Polar effective temperature (in K). 
(4) Equatorial effective temperature (in K), (5) Polar radius (in unit of the solar radius $R_{\odot}$).
(6) Equatorial radius (in $R_{\odot}$).(7) Rotational velocity at the equator (in km\,s$^{-1}$). 
(8) Inclination angle of the rotational axis (in degree). (9) Projected rotational velocity (in km\,s$^{-1}$). 
(10) Gravitational darkening parameter ($T_{\rm eff} \propto g^{\beta}$). (11) Specific remark. 
At the last row, the input parameters ($M$, $T_{\rm eff,p}$, $R_{\rm p}$, $v_{\rm e}$, $i$, and $\beta$; 
taken from Monnier et al.) along with the output results ($T_{\rm eff,e}$ and $R_{\rm e}$) of 
the SEDINT program are also presented.
\label{tab1}
\end{table*}

In contrast, regarding the surface abundances of Altair, an important 
topic in connection with chemical peculiarities often seen in A-type stars, 
little is known unfortunately, since trials of elemental abundance 
determinations for this star have been scarce. To the author's 
knowledge, two spectroscopic studies have worked upon Altair's 
chemical abundances of various elements (though the global metallicity 
was reported in some other work; cf. Table~\ref{tab2}). That is, \citet{Erspamer+North:2003} 
included  Altair in their abundance studies for 140 A--F stars.
In addition, Altair was also included in \citet{Luck:2017}'s recent 
spectroscopic study for parameters and abundances of 1002 stars 
in the local region, though his project was essentially focused on 
late-type stars (F, G, and K dwarfs and giants).

However, their consequences are not consistent with each other. For example,
regarding the microturbulence ($v_{\rm t}$) and Fe abundance ([Fe/H]),
\citet{Erspamer+North:2003} derived 2.0\,km\,s$^{-1}$ and $-0.24$\,dex,
while appreciably larger values of 5.7\,km\,s$^{-1}$ and $+0.48$\,dex
resulted from \citet{Luck:2017}'s analysis. Besides, abundance patterns
are apparently different. \citet{Luck:2017}'s results show considerable
overabundances ([X/H]\,$\gtrsim 0.5$; some are even 
$1 \lesssim$\,[X/H]\,$\lesssim 2$) in heavier elements (Ti, V, Cr, Mn, Fe, 
Co, Ni, Y, Zr, Ce, Nd, Eu, except for Ba which is underabundant), but Ca 
and Sc are inversely underabundant ([X/H]\,$\lesssim -0.5$), interestingly 
suggesting characteristic peculiarities seen in A-type metallic-line 
(Am) stars. On the other hand, such Am-like trends are not observed in 
the [X/H] values derived by \citet{Erspamer+North:2003}, which tend to be
within  $-0.3 \lesssim$\,[X/H]\,$\lesssim +0.3$ for many elements,
excepting that overabundances of Na, Sc, Co, La, Nd by $\sim$\,+0.5--0.9~dex
while a large underabundance of Sr by $\sim -2$\,dex.  

Presumably, such a paucity of chemical abundance studies and diversified results 
for Altair are due to the fact that reliable abundance determinations
are very difficult for such a conspicuously rapid rotator 
($v_{\rm e}\sin i \sim$\,200--250\,km\,s$^{-1}$), in which spectral lines  
are considerably broadened and blurred out, causing serious blending 
with each other. Actually, chemical abundances of A-type stars with
$v_{\rm e}\sin i \gtrsim$\,200\,km\,s$^{-1}$ seem to have been rarely 
investigated so far. though trials for moderately fast rotators 
($100 < v_{\rm e}\sin i < 200$\,km\,s$^{-1}$) turned out somehow 
feasible if carefully done \citep[e.g.,][]{Lemke:1993}.

In this investigation, we try to challenge this hard task of
establishing the photospheric elemental abundances of Altair.
The aim is to clarify whether or not this very rapidly-rotating A-type star 
in the solar neighborhood shows any chemical peculiarities.

It has been generally believed that the key parameter for the emergence of 
Am-like chemical peculiarities (CP) in A stars (which may possibly be related 
to element diffusion process requiring the stability of the atmosphere/envelope) 
is the rotational velocity, in the sense that the anomaly emerges for slowly 
rotating stars  while it is suppressed for the case of rapid rotation.
However, the critical rotational velocity demarcating CP and non-CP 
groups is not yet well understood. While \citet{Abt+Morrell:1995} stated that
A5--F0 stars with $v_{\rm e}\sin i > 120$\,km\,s$^{-1}$ are normal stars
(cf. their fig.~6), some chemical anomaly in C and Ba might still persist up 
to $v_{\rm e} \sin i \lesssim 180$\,km\,s$^{-1}$ as seen from \citet{Lemke:1993}'s 
fig.~3. Therefore, it is interesting to examine whether or not any Am-like 
abundance peculiarities (deficiencies of light elements such as C, N, and O
as well as Sc or Ca; overabundances of heavy elements such as s-process ones) 
are observed in Altair, an intrinsically very rapid rotator with 
$v_{\rm e}$ as large as $\sim 300$\,km\,s$^{-1}$.   

Alternatively,  there is another group of chemically peculiar stars 
(so-called $\lambda$~Boo stars) which show deficiencies in refractory 
Fe group elements (presumably caused by dust--gas separation mechanism) 
while volatile elements (such as CNO) remain almost normal 
\citep[e.g.,][]{Venn+Lambert:1990}. Since most stars of this group are rapid 
rotators, it may be possible that Altair shows this kind of anomaly, since 
$K$-band excess indicating circumstellar dust emission is observed
\citep{Nunez_etal:2017}. Accordingly, abundances of volatile and 
refractory elements should be compared each other in order to see if 
there is any difference between these. 

Methodologically, since application of spectrum-synthesis technique
is mandatory for the present case, we extensively employ the 
spectrum-fitting code developed by \citet{Takeda:1995}. This flux-based method 
(using only arbitrarily-scaled spectrum) is advantageous, because it 
does not require any precise continuum normalization in advance, which is
difficult especially for rapid rotators. 

\section{Atmospheric model parameters}\label{sec2}

As usual, we employ the conventional plane-parallel model atmosphere for abundance
determination, which is characterized by $T_{\rm eff}$ (effective temperature) and 
$\log g$ (logarithmic surface gravity). However, since (i) these parameters are 
significantly latitude-dependent [$T_{\rm eff}(\theta)$, $\log g(\theta)$] over the 
stellar surface and (ii) how these surface inhomogeneities affect the observed stellar 
flux depends upon the aspect angle of rotation ($i$) in the present case, we have to 
find ``adequately averaged'' $\langle T_{\rm eff} \rangle$ and $\langle \log g\rangle$ 
of Altair. Here, we invoke a theoretical rotating star model in order to calculate  
the intensity distribution ($I$) at any point of the stellar disk. Then, 
$\langle T_{\rm eff} \rangle$ and $\langle \log g\rangle$ may be 
derived by averaging the local $T_{\rm eff}$ and $\log g$ over the visible disk 
while weighting them with the corresponding brightness. This is the approach 
which was adopted also by \citet{Takeda:2023a}.

For this purpose, the program SEDINT developed by \citet{Takeda_etal:2008} was used, 
which simulates the spectral energy distribution (SED) of a rotating star 
based on the Roche model.  That is, SEDINT calculates the local specific
intensity directed toward the observer $I_{\lambda}(\xi,\eta)$ at each disk 
point $(\xi,\eta)$ projected on the sky, from which the theoretical flux 
observed at the earth ($F_{\lambda}$) is derived by integration as  
\begin{equation}
F_{\lambda}  \equiv 
\left. \int\!\!\!\int_{{\rm disk}} \; I_{\lambda}(\xi,\eta) \; {\rm d}\xi {\rm d}\eta \middle / d^{2} \right.
\end{equation} 
where $d$ is the distance to Altair (5.14~pc).

The input parameters of SEDINT are $M$ (stellar mass), $T_{\rm eff,p}$ 
(effective temperature at the pole), $R_{\rm p}$ (polar radius), 
$v_{\rm e}$ (equatorial rotational velocity), $\beta$ (gravitational darkening 
parameter), and $i$ (inclination angle).
As to the local model atmospheres, solar-metallicity models were employed.
Regarding these six parameters, we decided to adopt the values derived by 
\citet{Monnier_etal:2007} (cf. the last row in Table~\ref{tab1}), because the resulting 
$F_{\lambda}$ satisfactorily matches the observed SED ($f_{\lambda}$) by this 
choice. This is demonstrated in Fig.~\ref{fig1}a (blue line), where the results 
corresponding to \citet{Hadjara_etal:2014}'s parameters ($T_{\rm eff,p}$ higher 
by $\sim$\,400--500\,K) are also shown for comparison (green line).

Then, we may define $\langle T_{\rm eff} \rangle$ (mean $T_{\rm eff}$ averaged over 
the disk) and $\langle \log g \rangle$ (mean $\log g$ averaged over the disk) as 
\begin{equation}
\langle T_{\rm eff} \rangle  \equiv  
\frac{ \int\!\!\!\int_{{\rm disk}} \; T_{\rm eff}(\xi,\eta) \; I_{5000}(\xi,\eta) \; {\rm d}\xi {\rm d}\eta }  
 {\int\!\!\!\int_{{\rm disk}} \; I_{5000}(\xi,\eta) \; {\rm d}\xi {\rm d}\eta }
\end{equation}
and 
\begin{equation}
\langle \log g \rangle  \equiv 
\frac{ \int\!\!\!\int_{{\rm disk}} \; \log g(\xi,\eta) \; I_{5000}(\xi,\eta) \; 
{\rm d}\xi {\rm d}\eta }
 {\int\!\!\!\int_{{\rm disk}}\; I_{5000}(\xi,\eta) \; {\rm d}\xi {\rm d}\eta }, 
\end{equation}
where $I_{5000}$ is the specific intensity at 5000\,\AA.

The values of $\langle T_{\rm eff} \rangle$ and $\langle \log g \rangle$
resulting from Eqs.~(2) and (3) are 7663~K and 4.04. Accordingly, we adopt 
$T_{\rm eff} = 7660$~K (rounded value) and $\log g = 4.04$ as the standard 
parameters of Altair, and the corresponding \citet{Kurucz:1993}'s ATLAS9 
solar-metallicity model is used for abundance determination. 
The flux calculated with this model by using the ATLAS9 program agrees well 
with the observed SED as shown in Fig.~\ref{fig1}b. Such established $T_{\rm eff}$ 
and $\log g$ are also compared with various literature values in Table~\ref{tab2}. 

\setcounter{table}{1}
\begin{table*}[h]
\caption{Conventional atmospheric parameters of Altair adopted in past publications.}
\scriptsize
\begin{center}
\begin{tabular}{cccccl} 
\hline\hline
Literature & $T_{\rm eff}$ & $\log g$ & $v_{\rm t}$ & [Fe/H] & Remark \\
(1)    &   (2)  & (3)  & (4) & (5) & (6) \\
\hline
\citet{Sokolov:1995}        &  8040 &      &      &        & Use of Balmer continuum slope \\
\citet{DiBenedetto:1998}   &  7568 &      &      &        & Use of $V-K$ color  \\
\citet{Erspamer+North:2003} & 7550 & 4.13 & 2.0 & $-0.24$ & Chemical abundance analysis \\
\citet{Gray_etal:2003}    &  7800 & 3.76 &  2.0 & +0.02$^{*}$  & Fitting with synthetic spectrum  \\
\citet{AllendePrieto_etal:2004} & 7646 & 4.23 &    &   &  Photometric $T_{\rm eff}$, $\log g$ from isochrones  \\
\citet{Zorec+Royer:2012} &  7727 &      &      &        & Use of $uvby\beta$ color  \\
\citet{Luck:2017}           &  7377 & 3.95 &  5.7 & +0.48  & Chemical abundance analysis \\
\citet{Borisov_etal:2023} &  7485 & 3.72 &    &  $-0.08^{*}$ & Fitting with synthetic spectrum grid \\
\hline
This study            &  7660 & 4.04 &  2.9 & $-0.12$ &   \\
\hline
\end{tabular}
\end{center}
(1) Reference. (2) Effective temperature (in K). (3) Logarithmic surface gravity (in c.g.s unit).
(4) Microturbulent velocity dispersion (in km\,s$^{-1}$).(5) Fe abundance relative to the Sun (in dex).
(6) Specific remark. At the last row, the parameters derived/adopted in this study are also given
for comparison.\\ 
$^{*}$This value corresponds to [M/H] (metallicity; i.e., logarithmic scale factor common to all 
elements applied to solar compositions).
\label{tab2}
\end{table*}

\setcounter{figure}{0}
\begin{figure}[h]
\begin{center}
\includegraphics[width=6.0cm]{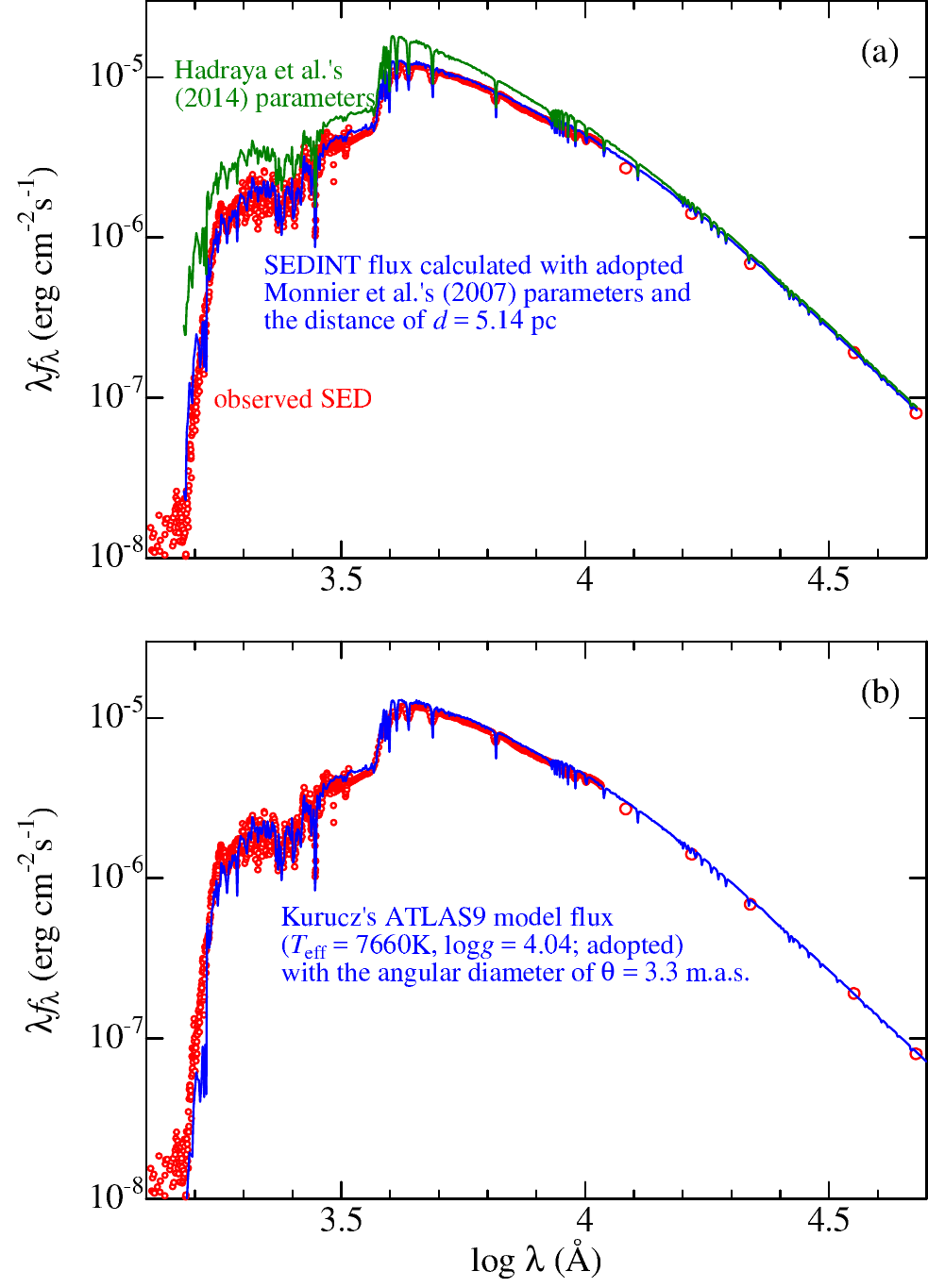}
\caption{
Theoretically calculated spectral energy distributions (SED) (depicted 
in lines) are compared with the observed SED of Altair (red symbols).
The observational data were taken from the archive data of International 
Ultraviolet Explorer (SWP48485LL+LWP15801LL) in the UV region, 
\citet{Alekseeva_etal:1996} in the visible region, and 
\citet{Bouchet_etal:1991} with \citet{Cohen_etal:1992}'s
calibration in the infrared region ($JHKLM$ band).
(a) Theoretical SED was calculated from the gravity-darkened rotating
star model by using the SEDINT program \citep{Takeda_etal:2008} by using 
\citet{Monnier_etal:2007}'s parameters we adopted (blue lines). In addition,
the result with \citet{Hadjara_etal:2014}'s parameters (not adopted) is 
also depicted in green lines for reference.
(b) Theoretical SED (blue lines) is the flux calculated from \citet{Kurucz:1993}'s 
ATLAS9 solar-metallicity model corresponding to ($T_{\rm eff}$, $\log g$) = 
(7660~K, 4.04) adopted in this study. 
}
\label{fig1}
\end{center}
\end{figure}

\section{Method and data for the analysis}\label{sec3}

\subsection{Spectrum fitting method}\label{sec3.1}

In this investigation, we extensively employ the stellar spectrum analysis 
tool MPFIT \citep{Takeda:1995}, which was developed based on \citet{Kurucz:1993}'s 
ATLAS9/WIDTH9 program and has a function of establishing the spectrum-related 
parameters (elemental abundances, microturbulence, macrobroadening parameters, 
radial velocity shift, etc.) by automatically searching for the best-fit solutions 
without any necessity of precisely placing the continuum level in advance.

In applying this method to the spectra of Altair, where shallow and heavily blended 
line features distribute over wide wavelength ranges (several tens of \AA), it is essential 
to introduce not only the scale-controlling constant ($C$) but also the tilt-adjustment 
parameter ($\alpha$) in comparing the observed flux ($f_{\lambda}$; in arbitrary scale) 
with the theoretical flux ($F_{\lambda}$) in order to accomplish a satisfactory fit. 
That is,  the right-hand side of equation (1) in \citet{Takeda:1995} is redefined as 
$\sum^{}_{} \{\log f_{\lambda} - \log F_{\lambda} - C - \alpha 
(\lambda - \lambda_{\rm min})/(\lambda_{\rm max}- \lambda_{\rm min})\} / N$,
and the best values of $C$ and $\alpha$ are iteratively established, 
where $\lambda_{\rm max}$ and  $\lambda_{\rm min}$ are the maximum and minimum 
wavelength of the relevant region.

In the present case, the parameters to be varied to achieve the best fit are (i) the 
abundances of selected $N$ elements ($A_{1}$, $A_{2}$, $\ldots$, $A_{N}$; cf. Table~\ref{tab3}), 
(ii) the projected rotational velocity ($v_{\rm e}\sin i$), and (iii) the wavelength shift 
($\Delta \lambda_{\rm r}$).The abundances of all other elements remain unchanged
at the solar abundances (i.e., model atmosphere values), and the microturbulence 
($v_{\rm t}$) is fixed (which is determined in Sect.~\ref{sec4.1}). 
The assumption of LTE (Local Thermodynamic Equilibrium) is postulated at this stage 
of spectrum fitting analysis. 

\subsection{Atomic line data}\label{sec3.2}

Regarding the data (wavelength, excitation potential, $gf$ value, damping parameters) 
of spectral lines in the relevant wavelength regions (3900--9300\,\AA), we exclusively 
consulted Vienna Atomic Line Database\footnote{https://vald.astro.uu.se/} 
\citep{Ryabchikova_etal:2015}. 
Because the data were downloaded about 10 years ago (2016 July) 
and not the latest, some modifications were further applied.
(1) As the $gf$ values of a number of neutral carbon lines turned out to be too large 
in the old data, all the data of C~{\sc i} lines were replaced by those newly downloaded
in 2026 January. (2)  Since the $gf$ values of two Si~{\sc i} lines at 6353.3597 and
6353.5188\,\AA\ were found to be unreasonably large, these two lines were excluded. 

\subsection{Observational data}\label{sec3.3}

As to the observed spectra of Altair, the data published by \citet{AllendePrieto_etal:2004}
were basically employed in this study, which are the data resulting from the project 
``Spectroscopic Survey of Stars in the Solar Neighborhood'' 
(abbreviated as S$^{4}$N).\footnote{https://hebe.as.utexas.edu/s4n/}
These spectra (with a resolving power of $R \sim 50000$) have sufficiently high 
signal-to-noise ratio of several hundred and cover the wavelength range of 3630--10440\,\AA.

In addition, the spectra of the UVES Paranal Observatory Project 
(UVES-POP)\footnote{https://www.eso.org/sci/observing/tools/uvespop.html} 
\citep{Bagnulo_etal:2003} were also used for several specific regions,
because S$^{4}$N spectra have some spike-like feature ($\lambda$\,4465--4495\,\AA\ region) 
or telluric lines are less severe for the case of the UVES-POP spectra 
($\lambda > 9000$\,\AA\ region).
 
When necessary, telluric lines were removed by hand-drawing the upper envelope 
for five conspicuously contaminated spectral regions, as depicted in Fig.~\ref{fig2}.  

\setcounter{figure}{1}
\begin{figure}[h]
\begin{center}
\includegraphics[width=6.0cm]{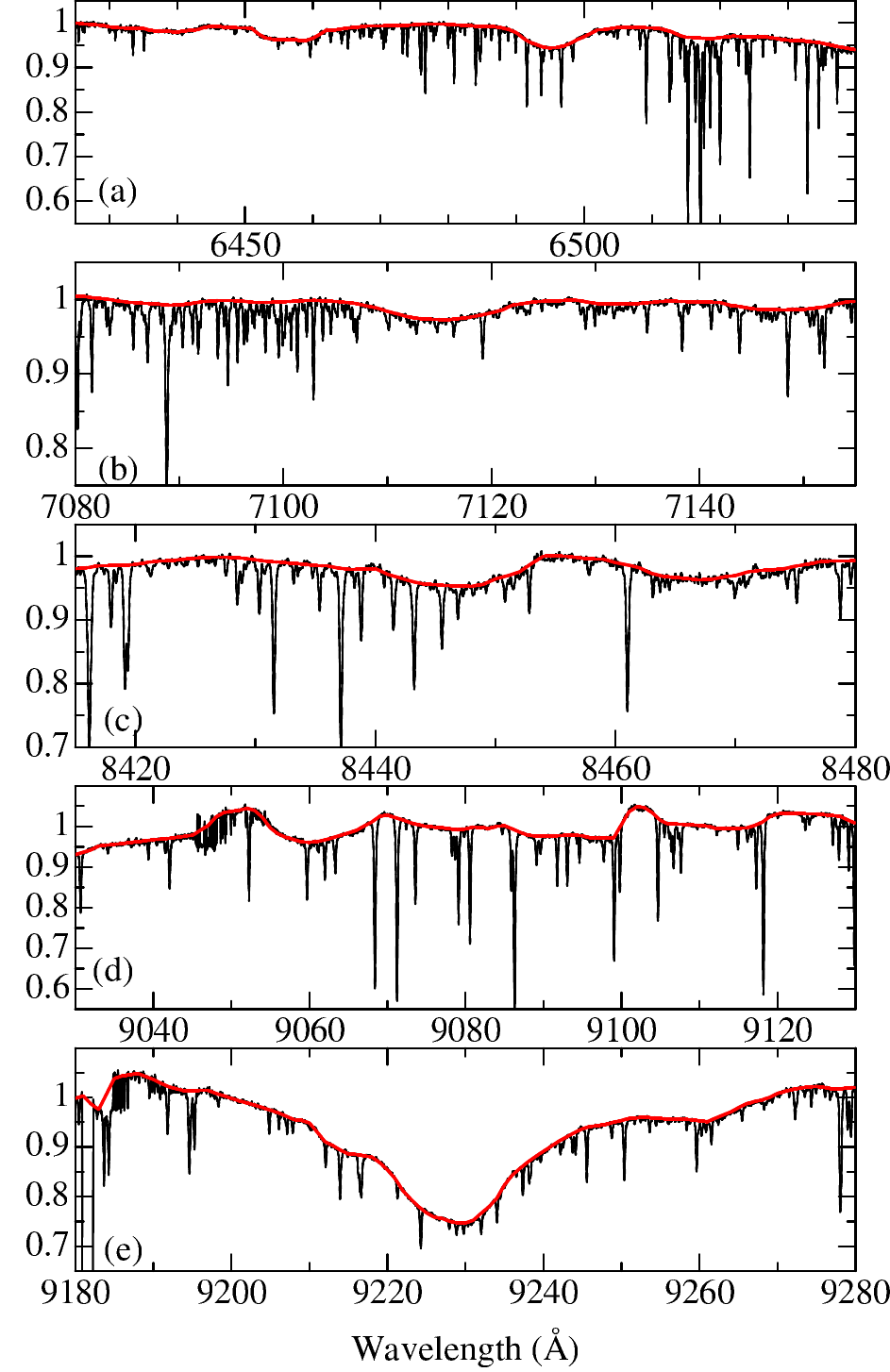}
\caption{
How the telluric lines were eliminated by hand-drawing the upper 
envelope (red) of the real spectrum (black) is shown in five regions:
(a) 6425--6540\,\AA\ region, (b) 7080--7155\,\AA\ region,
(c) 8415--8480\,\AA\ region, (d) 9030--9130\,\AA\ region,
and (e) 9180--9280\,\AA\ region.
}
\label{fig2}
\end{center}
\end{figure}

\section{Determination of elemental abundances}\label{sec4}

\subsection{Microturbulence}\label{sec4.1}

Before deriving chemical abundances, we have to evaluate the microturbulent
velocity, an important parameter for abundance determination.
Usually, this parameter is determined by requiring that the abundances
derived from various lines (Fe lines in most cases) do not show any systematic 
strength-dependent trend. 

Since heavily blended spectra are concerned here and accuracy or difficulty in 
deriving the abundance for a specific element (e.g., Fe) differs from case to case, 
we follow a somewhat modified approach while assuming that elemental abundances 
are controlled by a single scaling parameter ([M/H]; metallicity) as
$A_{*,i}$ = $A_{\odot,i}$ + [M/H] for all elements $i$ \citep[such as done 
by ][]{Gray_etal:2003}, which is not a bad approximation for the present case of 
Altair (as we will see in Sect.~\ref{sec5.1}). Then, the abundance parameter 
to be varied is only [M/H].

In order for a successful determination of $v_{\rm t}$, it is necessary to derive [M/H]
from wavelength regions containing strong lines (quite sensitive to changing $v_{\rm t}$) 
as well as from those with weaker lines (only weakly $v_{\rm t}$-dependent).
Since lines are generally strong and numerous at shorter wavelengths (e.g., violet 
region) while they get weaker and fewer at longer wavelengths (e.g., green--yellow 
region), we selected 56 spectral regions (each being 30\,\AA-wide), where their central 
wavelengths are changed from 3900\,\AA\ to 5000\,\AA\ with a step of 20\,\AA\ (i.e.,
overlapping of 5\,\AA\ with each other). The spectrum-fitting technique was then applied 
to each of the 56 regions to derive [M/H] while changing $v_{\rm t}$ from 0.5\,km\,s$^{-1}$
to 5.0\,km\,s$^{-1}$ with a step of 0.5 \,km\,s$^{-1}$. 

Although not all trials were successful (e.g., iterative solutions were not converged 
after all or the behavior of [M/H] in response to changing $v_{\rm t}$ was anomalous), 
reasonable results were finally obtained for 40 regions. How the theoretical spectrum 
fits well with the observed one in each region is shown in Fig.~\ref{fig3}.    
The resulting [M/H] vs. $v_{\rm t}$ relations are depicted in Fig.~\ref{fig4}a,
and the mean $\langle$[M/H]$\rangle$ averaged over 40 regions as well as the 
corresponding standard deviation $\sigma$ are plotted against $v_{\rm t}$ in 
Figs.~4b and 4c, respectively. Fig.~\ref{fig4}c suggests that the microturbulence of Altair 
(corresponding to the minimum of $\sigma$) is $v_{\rm t}$ = 2.9\,($\pm 0.9$)\,km\,s$^{-1}$,
where the probable uncertainty was estimated as done in sect.~3.2 of \citet{Takeda:2022a}.

\setcounter{figure}{2}
\begin{figure*}[h]
\begin{center}
\includegraphics[width=14.0cm]{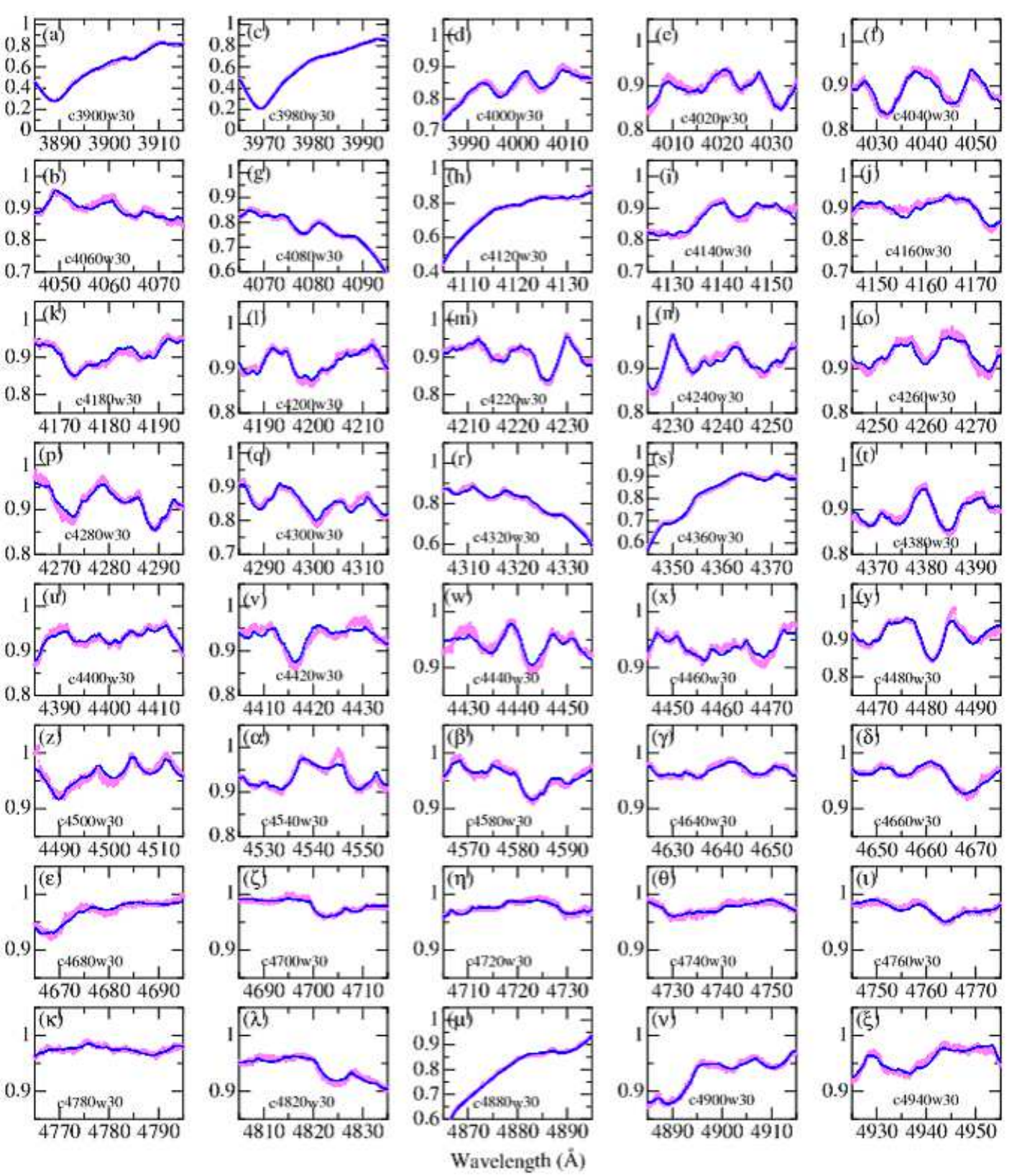}
\caption{
Fitting of the theoretical spectrum (blue lines) with the 
observed spectrum (pink symbols) by adjusting [M/H] at each of the 
40 regions (the region code is indicated in each panel, where ``c????w**''
means that the center/width of the region is ????\AA/**\AA)
for the purpose of $v_{\rm t}$ determination. The wavelength scale 
of the spectrum is adjusted to the laboratory frame, and the 
scale marked in the left ordinate corresponds to the theoretical 
residual flux ($F^{\rm th}_{\lambda}/F^{\rm th}_{\rm cont}$)
}
\label{fig3}
\end{center}
\end{figure*}

\setcounter{figure}{3}
\begin{figure}[h]
\begin{center}
\includegraphics[width=5.0cm]{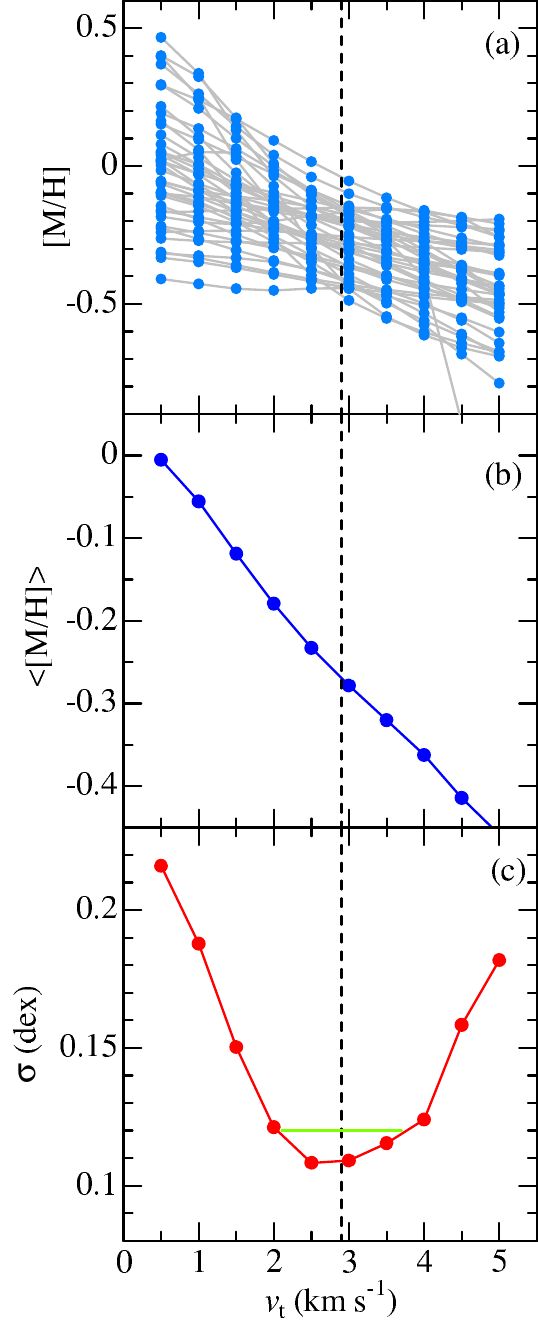}
\caption{
(a) [M/H] vs. $v_{\rm t}$ diagram constructed from the [M/H] 
results of 40 regions derived with 10 $v_{\rm t}$ values 
(0.5, 1.0, 1.5, $\cdots$, 4.5, and 5.0\,km\,s$^{-1}$).
(b) $\langle$[M/H]$\rangle$ (mean [M/H] averaged for all 40 regions)
plotted against $v_{\rm t}$.
(c) $\sigma$ (standard deviation of [M/H]) plotted against $v_{\rm t}$.
The green horizontal segment denotes the possible uncertainty of $v_{\rm t}$
($\pm 0.9$\,km\,$^{-1}$) around the minimum-$\sigma$ solution (2.9\,km\,s$^{-1}$),
which was estimated from the fluctuation of $\sigma$ based on the numerical 
simulation \citep[cf. sect.~3.2 in ][]{Takeda:2022a}. At each panel, the position of 
$v_{\rm t}$ = 2.9\,km\,$^{-1}$ is indicated by the vertical dashed line.
}
\label{fig4}
\end{center}
\end{figure}

\subsection{Abundances derived by spectrum fitting}\label{sec4.2}

Now that $v_{\rm t}$ has been established, the next task is to determine the
chemical abundances of various elements. Given the motivations described
in Sect.~\ref{sec1}, special attention is paid to the following elements: 
(i) those representing typical Am peculiarities (underabundances in C, N, O, Ca, 
and Sc; overabundances in heavier species such as Fe group elements as well as 
s-process elements like Sr or Ba). (ii) those of volatile species (C, N, O, S, and Zn) 
which remain near-normal in $\lambda$~Boo stars in contrast to other refractory 
species (such as Fe group) showing deficiencies. 

After preparatory simulations of theoretical line strengths over wide wavelength 
ranges, followed by test trials of spectrum fitting to check the feasibility, 
a total of 28 regions (typically several tens of \AA\ wide) were eventually 
selected and confirmed to work out. These spectral regions along with the 
target elements (their abundances are varied to accomplish the best fit) are 
summarized in Table~\ref{tab3}, which indicates that the abundances of 17 elements 
(C, N, O, Mg, Al, Si, S, Ca, Sc, Ti, Cr, Mn, Fe, Ni, Zn, Sr, and Ba) were determined. 
The theoretical spectra corresponding to the converged solutions are compared  
with the observed spectra in Fig.~\ref{fig5} for each region.

\setcounter{table}{2}
\setlength{\tabcolsep}{3pt}
\begin{table}[h]
\scriptsize
\caption{Selected spectrum regions and target elements.}
\begin{tabular}{cc}\hline\hline
Region code & Elements \\
(1)         & (2)       \\
\hline
39253950 & Ca, Al, Fe             \\ 
40604090 & Fe, Sr, Ni             \\
42004230 & Ca, Sr, Fe             \\
42354265 & Fe, Sc, Cr, Mn         \\
42954315 & Fe, Ti, Ca, Sc         \\
44654495 & Mg, Ti, Fe, Ni         \\
45404570 & Ti, Fe, Cr, Ba         \\
46604690 & Fe, Sc, Ni, Zn         \\
47104745 & Fe, Ni, Zn, Cr         \\
48004830 & Cr, Ti, Mn, Zn, Ni, Fe \\
50155045 & Fe, Ca, Ni, Sc, Si     \\
50355070 & Fe, Ni, C, Si, Ca      \\
51605200 & Mg, Fe, Ti             \\
53605400 & Fe, C, Ti              \\
60306060 & Fe, Si, S              \\
61206150 & Ca, Ba, Fe             \\
61506180 & Ca, Si, O, Fe, Ni      \\
63406380 & Si, Ca                 \\
64806510 & Fe, Ca, Ba, Ti         \\
67256765 & S, Fe                  \\
71007140 & Ni, C, Fe              \\
74357475 & Fe, Cr, N              \\
77607790 & O, Fe                  \\
84258460 & O. Fe, Si              \\
86758705 & Fe, N, S               \\
90509080 & C                      \\
92009240 & S, Mg                  \\
92409280 & Mg, O                  \\
\hline
\hline
\end{tabular}
(1) Region code of 8 characters, where ``sssseeee'' indicates that the fitting 
analysis was done in the spectrum range from ssss\,\AA\ to eeee\,\AA. 
(2) Elements whose abundances were varied to accomplish the best fit, 
while those of other elements were fixed at the solar composition.
\label{tab3}
\end{table}

\setcounter{figure}{4}
\begin{figure*}[h]
\begin{center}
\includegraphics[width=14.0cm]{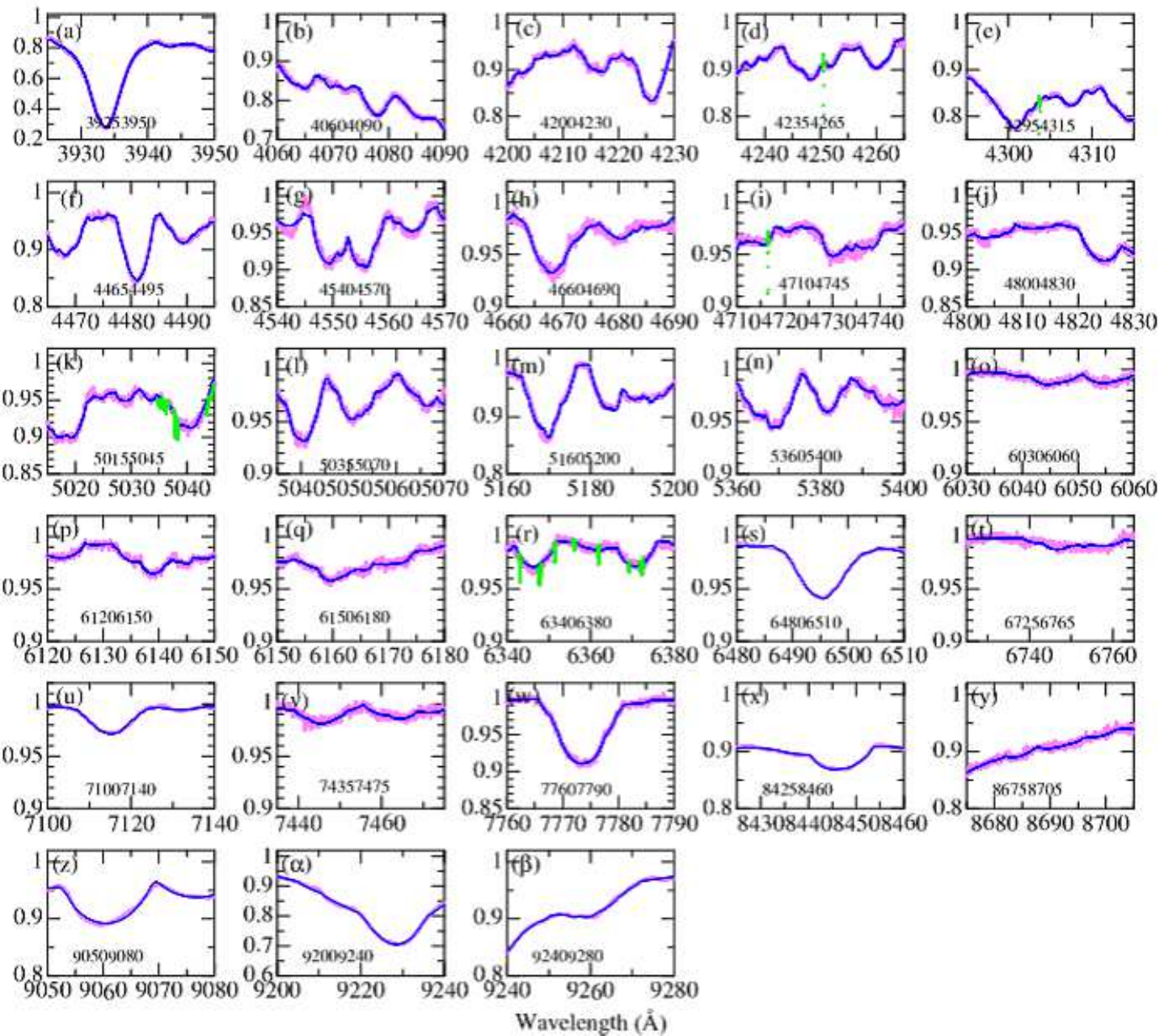}
\caption{
Fitting of the theoretical spectrum (blue lines) with the 
observed spectrum (pink symbols) at each of the 28 regions 
(the region code is indicated in each panel) for the purpose 
of determining the chemical abundances of various elements.
The masked portions (spectrum defect or telluric feature), 
which were excluded from judging the goodness of fit, are 
highlighted in light green. Otherwise, the same as in Fig.~\ref{fig3}.
}
\label{fig5}
\end{center}
\end{figure*}

\subsection{Application of non-LTE corrections}\label{sec4.3}

The LTE abundances derived in Sect.~\ref{sec4.2} should then be non-LTE corrected wherever possible.
Since multiple lines are generally involved with the fitting-based abundance of any element
resulting from a specific region, the non-LTE correction to be applied is evaluated as follows.
\begin{itemize}
\item
Let us consider an element whose LTE abundance was derived to be $A$ from the spectrum-fitting
analysis in a given region.
\item
First, the values of $\eta_{0}$ ($\equiv l_{0}/\kappa$; line-center to continuum opacity ratio 
at $\tau_{5000} \sim 0.2$) are evaluated by using this $A$ for all available lines of this 
element included in this region.   
\item
Then, we select only $K$ lines of appreciable strengths out of them, according to the criterion
of satisfying the condition $\eta_{0} > 0.1$; and their theoretical equivalent widths are 
calculated as ($W_{i}$, $i = 1, 2,\ldots , K$).
\item
For these $K$ lines, the non-LTE corrections corresponding to each $W_{i}$.are are computed
as ($\Delta_{i}$, $i = 1, 2,\ldots , K$).
\item
Now, the mean non-LTE correction ($\langle \Delta \rangle$) is defined as the weighted 
average of $\Delta_{i}$  
($\langle \Delta \rangle \equiv \sum_{i = 1}^{K}{W_{i}\Delta_{i}}/\sum_{i = 1}^{K}{W_{i}}$),
from which the non-LTE abundance is finally obtained as $A + \langle \Delta \rangle$.
\end{itemize}

\setcounter{table}{3}
\begin{table}[h]
\caption{References of non-LTE calculations.}
\scriptsize
\begin{tabular}{ccl}
\hline\hline
Elem. & $Z$ & References \\
(1)   & (2)      & (3)   \\
\hline
 C    &  6 & \citet{Takeda:1992} \\
 N    &  7 & \citet{Takeda:1992} \\
 O    &  8 & \citet{Takeda:2003} \\
 Mg   & 12 & \citet{Takeda:2025} \\
 Al   & 13 & \citet{Takeda:2023b} \\
 Si   & 14 & \citet{Takeda:2022b} \\
 S    & 16 & \citet{Takeda_etal:2005} \\
 Ca   & 20 & \citet{Takeda:2020} \\
 Sc   & 21 & (see Sect.~\ref{sec4.3}) \\
 Zn   & 30 & \citet{Takeda_etal:2005} \\
 Sr   & 38 & (see Sect.~\ref{sec4.3}) \\
 Ba   & 56 & \citet{Takeda:2026} \\
\hline
\end{tabular}
(1) Element. (2) Atomic number. (3) These papers (and the references quoted therein) may 
be consulted for more details about the calculations (e.g., adopted model atoms).
The non-LTE calculations for these elements were done by assuming the solar abundances.
This is not necessarily a bad approximation in the present case, and the resulting
departure coefficients are anyhow not much sensitive to a choice of input abundances. 
\label{tab4}
\end{table}

These non-LTE corrections were calculated for 12 elements (C, N, O, Mg, Al, Si,
S, Ca, Sc, Zn, Sr, and Ba) for which the author already has experiences 
of non-LTE calculations (cf. Table~\ref{tab4}), while LTE was assumed ($\langle \Delta \rangle = 0$) 
for the other 5 elements (Ti, Cr, Mn, Fe, and Ni). Note that the atomic models of 
Sc ($Z = 21$) and Sr ($Z = 38$) were newly constructed for the purpose of this study 
as described below.
\begin{itemize}
\item[$\circ$]
The non-LTE calculations for Sc~{\sc ii} were carried out by using the Sc~{\sc ii} 
model atom comprising 89 terms (up to 3d\,6f\;$^{1}$H at 91197 cm$^{-1}$) and 1402 
radiative transitions, which is based on the atomic data of \citet{Kurucz+Bell:1995}.
Though the contribution of Sc~{\sc i} was neglected, the levels of Sc~{\sc iii} 
were taken into account in the number conservation of total Sc atoms. 
The hydrogenic approximation was assumed for calculating photoionization rates.
Otherwise (such as the treatment of collisional rates), the procedure described 
in sect.~3.1.3 of \citet{Takeda:1991} was followed.  
\item[$\circ$]
Likewise, the Sr~{\sc ii} model atom was constructed from the atomic data of
\citet{Kurucz+Bell:1995}, which comprises 32 terms (up to 9d\;$^{2}$D 
at 81249 cm$^{-1}$) and 54 radiative transitions. Again, the contribution 
of Sr~{\sc i} was neglected, while the levels of Sr~{\sc iii} 
were taken into account in the number conservation of total Sr atoms. 
Regarding the photoionization cross section, the data read from fig.~1 of 
\citet{Mashonkina_etal:2020} were used for the lowest 3 terms (5s\,$^{2}$S, 4d\,$^{2}$D, 
and 5p\,$^{2}$P$^{\circ}$), while the hydrogenic approximation was assumed for 
the remaining terms. Collisional rates were calculated by following recipe 
in sect.~3.1.3 of \citet{Takeda:1991}.
\end{itemize}

The detailed results are presented in the online supplementary materials:
The non-LTE corrections ($\Delta_{i}$) of individual lines
as well as the resulting mean correction ($\langle \Delta \rangle$) 
for each element/region (along with relevant atomic line data) are given 
in ``nltelines.dat''. Meanwhile,  the region-by-region LTE/NLTE abundances
and the corresponding NLTE corrections for each element are summarized 
in ``reg\_abund.dat''.

\section{Examination of abundance results}\label{sec5}

\subsection{Mean abundance and standard deviation}\label{sec5.1}

Based on the final abundances of 17 elements derived in Sect.~\ref{sec4.3}, 
their mean values ($\langle A_{*}^{\rm X}\rangle$) averaged over $N$ available
regions are summarized in Table~\ref{tab5}, where the reference solar abundances ($A_{\odot}^{\rm X}$), 
mean differential abundances relative to the Sun ($\langle$[X/H]$\rangle$), and the  
standard deviations around the mean ($\sigma$) are also presented. 
We consulted \citet{Anders+Grevesse:1989}'s compilation\footnote{Note that 
these data may be somewhat outdated especially for C, N, and O according to 
more recent compilations. For example, \citet{Asplund_etal:2009} derived (by taking 
into account the 3D effect) 8.43 (C), 7.83 (N), and 8.69 (O) for the solar photospheric 
abundances, which are by $\sim 0.2$\,dex lower than by the values we adopted.
For other elements, the differences between \citet{Anders+Grevesse:1989} and 
\citet{Asplund_etal:2009} are insignificant.} for $A_{\odot}$ 
(except for Fe for which $A_{\odot} = 7.50$ was adopted), in order to maintain 
consistency with \citet{Kurucz:1993}'s ATLAS9 models employed in this study. 
The resulting mean $\langle$[X/H]$\rangle$ and its standard
deviation $\sigma$ (along with individual region-by-region values of [X/H]) for each 
element are graphically depicted in Fig.~\ref{fig6}.

An inspection of Table~\ref{tab5} and Fig.~\ref{fig6} reveals that the mean abundances 
averaged over regions show considerably large $\sigma$ (up to $\sim$\,0.4--0.5\,dex 
or even more) for several elements. We examine below these problematic cases 
of $\sigma > 0.4$ (i.e., C, S, Cr, Ni, and Zn) somewhat more in detail.

\subsection{Carbon}\label{sec5.2}

The abundance of C from the 90509080 region (9.34) is apparently an outlier
compared to those (8.2--8.4) from other three regions (50355070, 53605400, 71007140).
That is, the C abundance derived from the strong 9061/9062/9078 lines of multiplet 1
may still be significantly overestimated despite of their large (negative) NLTE
corrections  ($\sim -0.5$\,dex). If the data from this region is neglected,
the mean abundance is reduced by 0.26\,dex down to $\langle$[C/H]$\rangle$ = $-0.27$

\subsection{Sulfur}\label{sec5.3}

Regarding S, the abundance from the 86758705 region (7.75) is markedly larger
than those (6.6--7.1) from other three regions (60306060, 67256765, 92009240).
That is, the S abundance derived from the 8693--8695 triplet lines 
might as well be overestimated. If the data from this region is neglected,
the mean abundance is reduced by 0.22\,dex down to $\langle$[S/H]$\rangle$ = $-0.35$

\subsection{Chromium}\label{sec5.4}

The Cr abundances derived from 5 regions widely range over $\sim$\,4.9--6.1,
among which that from the 45404570 region (4.87) is especially low. If we discard
this data, the mean increases by 0.20\,dex, yielding $\langle$[Cr/H]$\rangle$ = 
$+0.17$.

\subsection{Nickel}\label{sec5.5}

The abundances of Ni derived from 8 regions widely range over $\sim$\,5.5--7.0, 
among which that from the 46604690 region (7.01) is appreciably high compared to
the others. If this data is rejected, the mean decreases by 0.13\,dex, 
yielding $\langle$[Ni/H]$\rangle$ = $-0.32$. Since the number of available
regions is rather large ($N = 8$) in this case, the effect of excluding the outlier 
is comparatively less significant for this element.

\subsection{Zinc}\label{sec5.6}

Most serious is the case of Zn abundances, because the results (5.72, 4.81, and 
3.70) derived from the 46604690, 47104745, and 48004830 regions (each containing
Zn~{\sc i} 4680, 4722, and 4810 lines) extraordinarily differ from each other.
which results in a conspicuously large $\sigma$ of $\sim 0.8$\,dex.
Unfortunately, we can not judge which of the three may be reliable or unreliable 
in this case.

\subsection{Impact of excluding outlier data}\label{sec5.7}

We have thus evaluated how the mean abundances for C, S, Cr, and Ni vary 
if the data of appreciable deviations are excluded, which are on the order of 
$\sim 0.2$\,dex. These changes in $\langle$[X/H]$\rangle$ are also indicated
by red arrows in Fig.~\ref{fig6}. As seen from this figure, the impact of these variations
is not so significant as to essentially alter the global tendency of chemical abundances. 

\subsection{Effect of parameter changes}\label{sec5.8}

How the abundance results are affected by the atmospheric parameters was also 
examined by repeating the spectrum-fitting analysis while interchangeably varying 
$T_{\rm eff}$ by $\pm 200$\,K, $\log g$ by $\pm 0.2$\,dex, and $v_{\rm t}$ by 
$\pm 0.9$\,km\,s$^{-1}$. The amounts of variations turned out $|\delta_{T}| \lesssim 0.2$\,dex,   
$|\delta_{g}| \lesssim 0.1$\,dex, while $|\delta_{v}|$ is considerably diversified 
in the range of $\sim$\,0.0--0.5\,dex ($\sim$\,0.2\,dex on the average) because regions
containing strong saturated lines (e.g., Ba~{\sc ii} lines) are quite $v_{\rm t}$-sensitive.
These variations are not so important as to substantially affect the qualitative trends of
Fig.~\ref{fig6} (as was the case in Sect.~\ref{sec5.7}), 

\section{Discussion}\label{sec6}

\subsection{Chemical abundance trends of Altair}\label{sec6.1}

Given the substantial difficulty of abundance determination in such 
a very rapid rotator as Altair. the results are generally less reliable 
as compared to the case of slower rotators. Since a number of lines 
are intricately merged at any wavelength point under the large rotational 
broadening (at least over several \AA\ wide), only a slight spectrum defect 
or an imperfection in the data of atomic lines would easily lead to large 
errors or spurious consequences, which must be the reason for the considerable 
region-to-region abundance dispersion observed in several elements (cf. Sect.~\ref{sec5}).

Yet, based on the [X/H] results of 17 elements depicted in Fig.~\ref{fig6}, we may read 
the following characteristics regarding the photospheric abundances of Altair. 
\begin{itemize}
\item
Fig.~\ref{fig6} indicates that $\langle$[X/H]$\rangle$ ranges within 
$-0.5 \lesssim \langle$[X/H]$\rangle \lesssim +0.3$ for all 17 elements 
without showing any systematic dependence upon $Z$, which means that 
any characteristic abundance patterns of CP stars (cf. Sect.~\ref{sec1}) are absent. 
That is, the possibility of being an Am star (deficiency in light elements 
such as C, N, and O along with Ca/Sc; while overabundances of heavier elements 
such as Sr and Ba) is clearly excluded.  Likewise, specific characteristics seen 
in $\lambda$~Boo stars, in which volatile elements (C, N, O, S, Zn) are nearly 
normal whilst many other refractory species (such as Fe group ones) show 
appreciable deficiencies, are not observed either. Accordingly, we may state 
that Altair is not a chemically peculiar star but rather belongs to the 
group of normal stars.   
\item
However, the metallicity of this star is not necessarily the same as that of
the Sun in spite of its being in the solar neighborhood, which shows a somewhat
subsolar trend. That is, the average of $\langle$[X/H]$\rangle$ for 17 elements 
given in Table~\ref{tab5} is $-0.12$ (standard deviation is 0.22), which is further 
lowered to $-0.14$ if the outlier data are excluded for C, S, Cr and Ni 
(see the caption of Table~\ref{tab5}). This is almost consistent with the [M/H] value 
ranging from $\sim -0.2$ to $\sim -0.3$ (obtained as a by-product of 
$v_{\rm t}$-determination; cf. Fig.~\ref{fig4}b). Therefore, Altair appears to have a 
slightly lower metallicity in comparison with the Sun, though this should not 
be regarded as an acquired chemical anomaly but may rather be due to 
fluctuations in the galactic gas at the time of star formation. 
\end{itemize}

\setcounter{table}{4}
\begin{table}[h]
\caption{Elemental abundance results of Altair.}
\scriptsize
\begin{tabular}{cccccccc}
\hline\hline
X & $N$  & $\langle A^{\rm X}_{*}\rangle$ & $A^{\rm X}_{\odot}$ & $\langle$[X/H]$\rangle$ & 
$\sigma$ & [X/H]$_{\rm E}$ & [X/H]$_{\rm L}$\\
(1) & (2) & (3) & (4) & (5) & (6) & (7) & (8) \\
\hline
 C  &  4 & 8.55$^{*}$ & 8.56 & $-0.01^{*}$ & (0.46) & $-0.11$ & +0.11 \\
 N  &  2 & 7.79 & 8.05 & $-0.26$ & (0.01) &       &  \\
 O  &  4 & 8.70 & 8.93 & $-0.23$ & (0.16) & +0.32 & +0.07 \\
 Mg &  3 & 7.63 & 7.58 & $+0.05$ & (0.20) & +0.02 &  \\
 Al &  1 & 5.96 & 6.47 & $-0.51$ & (0.00) &       &  \\
 Si &  6 & 7.53 & 7.55 & $-0.02$ & (0.16) & +0.28 & +0.36 \\
 S  &  4 & 7.08$^{\dagger}$ & 7.21 & $-0.13^{\dagger}$ & (0.42) &       &  \\
 Ca &  9 & 6.14 & 6.36 & $-0.22$ & (0.23) & $-0.18$ & $-0.49$ \\
 Sc &  4 & 3.34 & 3.10 & $+0.24$ & (0.27) & +0.69 & $-1.28$ \\
 Ti &  7 & 5.03 & 4.99 & $+0.04$ & (0.36) & $-0.12$ & +1.19 \\
 Cr &  5 & 5.64$^{\ddagger}$ & 5.67 & $-0.03^{\ddagger}$ & (0.46) & $-0,14$ & +0.64 \\
 Mn &  2 & 4.97 & 5.39 & $-0.42$ & (0.23) & $-0.05$ & +1.27 \\
 Fe & 24 & 7.38 & 7.50 & $-0.12$ & (0.24) & $-0.24$ & +0.48 \\
 Ni &  8 & 6.06$^{\sharp}$ & 6.25 & $-0.19^{\sharp}$ & (0.44) & $-0.18$ & +1.19 \\
 Zn &  3 & 4.74 & 4.60 & $+0.14$ & (0.83) &       &  \\
 Sr &  2 & 2.39 & 2.90 & $-0.51$ & (0.16) & $-2.08$ &  \\
 Ba &  3 & 2.30 & 2.13 & $+0.17$ & (0.21) & +0.21 & $-0.62$ \\
\hline
\end{tabular}
(1) Element. (2) Number of regions. (3) Mean abundances averaged over $N$ regions.
(4) Reference solar abundances, which were taken from \citet{Anders+Grevesse:1989} 
except for Fe (see Sect.~\ref{sec5.1}). (5) Differential abundances relative
to the Sun defined as $\langle$[X/H]$\rangle$\,$\equiv \langle A^{\rm X}_{*} \rangle - A^{\rm X}_{\odot}$.
(6) Standard deviation of the mean abundance.
(7) [X/H] results of \citet{Erspamer+North:2003}.
(8) [X/H] results of \citet{Luck:2017}. \\
$^{*}$Reduced by $0.26$\,dex without 90509080 region data.\\
$^{\dagger}$Reduced by $0.22$\,dex without 86758705 region data.\\ 
$^{\ddagger}$Increased by $0.20$\,dex without 45404570 region data.\\
$^{\sharp}$Decreased by $0.13$\,dex without 46604690 region data.\\
\label{tab5}
\end{table}

\setcounter{figure}{5}
\begin{figure}[h]
\begin{center}
\includegraphics[width=6.0cm]{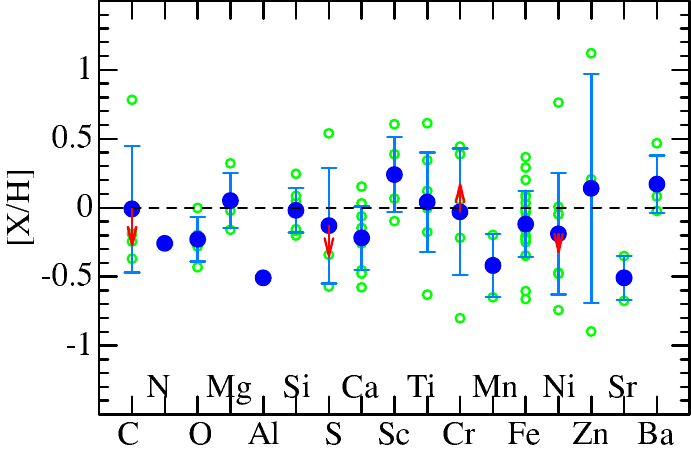}
\caption{
The [X/H] values (differential abundance of element X relative
to the Sun) of 17 elements resulting from the analysis of this study.
The mean values of $\langle$[X/H]$\rangle$ (averaged over available 
regions) are shown by blue bullets, where the error bars indicate 
the standard deviations. The individual [X/H] results of each spectrum 
region are also depicted by green open circles. The red arrows (shown 
in the plots of C, S, Cr, and Ni) indicate the amount/direction of 
changes if the outlier data were excluded (see  Sect.~\ref{sec5.7}).
}
\label{fig6}
\end{center}
\end{figure}

\subsection{Comparison with previous studies}\label{sec6.2}

It has thus been shown that Altair does not belong to the group of chemically 
peculiar stars (occasionally observed in late A- to late B-type stars of the upper 
main sequence), though its metallicity is slightly subsolar by $\sim$\,0.2\,dex. 
Let us examine how this consequence compares with the results reported in the 
past publications. Table~\ref{tab5} gives our [X/H] values (column~5) in comparison with 
those of \citet{Erspamer+North:2003} (column~7) and \citet{Luck:2017} (column~8).  

The results of \citet{Erspamer+North:2003} may be favorably compared with ours.
Actually, the difference ($|\Delta[{\rm X/H}]|$) is $\lesssim 0.3$\,dex for 
most cases, except for O (0.55\,dex), Sc (0.45\,dex), Mn (0.37\,dex), and Sr (1.57\,dex).
Therefore, although the discrepancy of Sr is extraordinarily large, their 
determinations are more or less tolerably consistent in view of the technical 
difference (e.g., their neglect of non-LTE effect) and inevitably lower precision 
of this kind of analysis. 

In contrast, serious inconsistencies are observed regarding the results of \citet{Luck:2017}.
While the differences are several tenths dex and not very conspicuous for lighter 
elements ($Z \le 20$), intolerably large discrepancies amounting to $\lesssim$\,1--2\,dex
are seen; e.g., Sc (1.52\,dex), Ti (1.15\,dex), Cr (0.67\,dex), Mn (1.69\,dex), 
Fe (0.60\,dex), Ni (1.38\,dex), and Ba (0.79\,dex). We should recall that \citet{Luck:2017}'s 
study was originally devoted to abundance determinations of FGK-type stars  and thus 
his analysis procedure (e.g., adopted lines) may not have been adequately adapted to 
A-type stars of higher $T_{\rm eff}$. Accordingly, we would not have to take this 
large discordance seriously.

\section{Summary and conclusion}\label{sec7}

Altair is a very rapidly-rotating A-type star, for which a number of researchers 
have investigated its physical parameters by way of direct interferometric observations 
(e.g., oblateness, rotation velocity, inclination angle, radius or $T_{\rm eff}$ 
at the pole/equator, gravitational darkening, etc.)

However, few spectroscopic studies on its photospheric chemical abundances have been 
conducted so far because of the considerable technical difficulty, which stems from the 
fact that spectral lines are widely broadened and blurred by rapid rotation and badly 
blended with each other. Moreover, the results of two available publications are 
apparently in conflict.
 
Motivated by this situation, we attempted a new spectroscopic analysis to establish 
the photospheric abundances of Altair while applying the synthetic spectrum-fitting 
technique to its high-dispersion spectra. Our aim was to clarify whether or not 
this A-type star shows any chemical peculiarities (like Am stars or $\lambda$~Boo stars). 

The microturbulent velocity was determined by requiring that the dispersion of 
[M/H] (metallicity) vs. $v_{\rm t}$ relation (derived for 40 regions selected 
from  violet--green wavelength ranges) be minimized, which yielded 
$v_{\rm t}=$\,2.9\,($\pm 0.9$)\,km\,s$^{-1}$.

The abundances of 17 elements (C, N, O, Mg, Al, Si, S, Ca, Sc, Ti, Cr, Mn, 
Fe, Ni, Zn, Sr, Ba) were then derived from the spectrum-fitting analysis applied 
to 28 selected regions, where the non-LTE effect was taken into consideration 
for 12 elements (C, N, O, Mg, Al, Si, S, Ca, Sc, Ti, Zn, Sr, and Ba)
while LTE was assumed for five Fe group elements (Ti, Cr, Mn, Fe, and Ni). 

The resulting abundances tend to show considerable region-by-region dispersion 
(amounting to several tenths dex or even more), which reflects the difficulty 
of reliable abundance determination for very rapid rotator, though the following 
conclusions could be somehow extracted regarding the chemical characteristics of Altair.

First, the region-averaged differential abundances relative to the Sun are within 
$-0.5 \lesssim$\,$\langle$[X/H]$\rangle$\,$\lesssim +0.3$ for all elements,
without showing any dependence upon $Z$ (or any meaningful difference between
volatile and refractory elements). Accordingly, we may rule out the possibility 
that Altair belongs to the group of chemically peculiar stars (such as Am stars 
or $\lambda$~Boo stars).

Second, since [M/H] (mean metallicity derived as a by-product of $v_{\rm t}$ 
determination) ranges from $\sim -0.2$ to $\sim -0.3$ and the mean 
$\langle$[X/H]$\rangle$ averaged over all elements is between $\sim -0.1$ to 
$\sim -0.2$, Altair's metallicity is slightly subsolar (by $\sim -0.2$\,dex on the average),
which is likely to be attributed to chemical fluctuations in the primordial gas.

\bmhead{Acknowledgements}

This research is based on the public-open data of the project ``Spectroscopic 
Survey of Stars in the Solar Neighborhood'' \citep{AllendePrieto_etal:2004}
as well as in part on those of the UVES Paranal Observatory Project \citep{Bagnulo_etal:2003}.   
This investigation has made use of the SIMBAD database, operated by CDS, 
Strasbourg, France, and the VALD database operated at Uppsala University,
the Institute of Astronomy RAS in Moscow, and the University of Vienna. 

\backmatter

\bmhead{Supplementary information}

The following online data are available as supplementary 
materials accompanied with this article. 
\begin{itemize}
\item
  ReadMe.txt 
\item
  nltelines.dat 
\item
  reg\_abund.dat
\end{itemize}

\section*{Declarations}

\begin{itemize}
\item Funding\\
The author declares that no funds, grants, or other support were received 
during the preparation of this manuscript.
\item Competing interests\\
The author has no relevant financial or non-financial interests to disclose.
\item Author contributions\\
This investigation has been conducted solely by the author.
\end{itemize}


\bibliography{Altair.bib}

\end{document}